\documentclass[aps,pra,twocolumn,showkeys,showpacs,unsortedaddress,bibnotes]{revtex4-1}

\usepackage{natbib,graphicx,dcolumn,color}
\usepackage{amsmath,amssymb,verbatim,bm}

\def\be{\begin{equation}}
\def\ee{\end{equation}}
\def\ber{\begin{eqnarray}}
\def\eer{\end{eqnarray}}

\def\sigmav{\mbox{\boldmath $\sigma$}}

\def\br{{\bf r}}
\def\bhz{{\bf \hat z}}
\def\bp{{\bf p}}

\def\bj{{\bf j}}

\def\Acalv{{\boldsymbol {\mathcal A}}}

\def\Fcalv{{\boldsymbol {\mathcal F}}}

\begin{document}

\title{Room-temperature transport properties of spin-orbit coupled Fermi systems:\\
       Spin thermoelectric effects, phonon skew scattering}

\author{Ulrich Eckern}
\affiliation{Institut f\"ur Physik, Universit\"at Augsburg, 86135 Augsburg, Germany}
\email{ulrich.eckern@physik.uni-augsburg.de}

\author{Cosimo Gorini}
\affiliation{Institut f\"ur Theoretische Physik, Universit\"at Regensburg, 93040 Regensburg, Germany}
\email{cosimo.gorini@physik.uni-regensburg.de}

\author{Roberto Raimondi}
\affiliation{Dipartimento di Matematica e Fisica, Universit{\`a} Roma Tre, 00146 Rome, Italy}
\email{roberto.raimondi@uniroma3.it}

\author{Sebastian T\"olle}
\affiliation{Institut f\"ur Physik, Universit\"at Augsburg, 86135 Augsburg, Germany}
\email{sebastian.toelle@physik.uni-augsburg.de}

\begin{abstract}
In this note
we summarize our recent results for the temperature dependence of transport coefficients
of metallic films in the presence of spin-orbit coupling. Our focus is on (i) the spin Nernst and the 
thermal Edelstein effects, and (ii) the phonon skew scattering contribution to the spin Hall conductivity,
which is relevant for the temperature dependence of the spin Hall angle. Depending on the parameters, the 
latter is expected to show a non-monotonous behavior.
\end{abstract}


\keywords{metallic films, spin-orbit interaction, kinetic theory, transport coefficients}
\pacs{72.25.−b, 72.10.Di, 72.15.Jf}

\maketitle

\section{Introduction}
\label{intro}

\begin{figure}[b]
\begin{center}
\includegraphics[width=0.8\columnwidth]{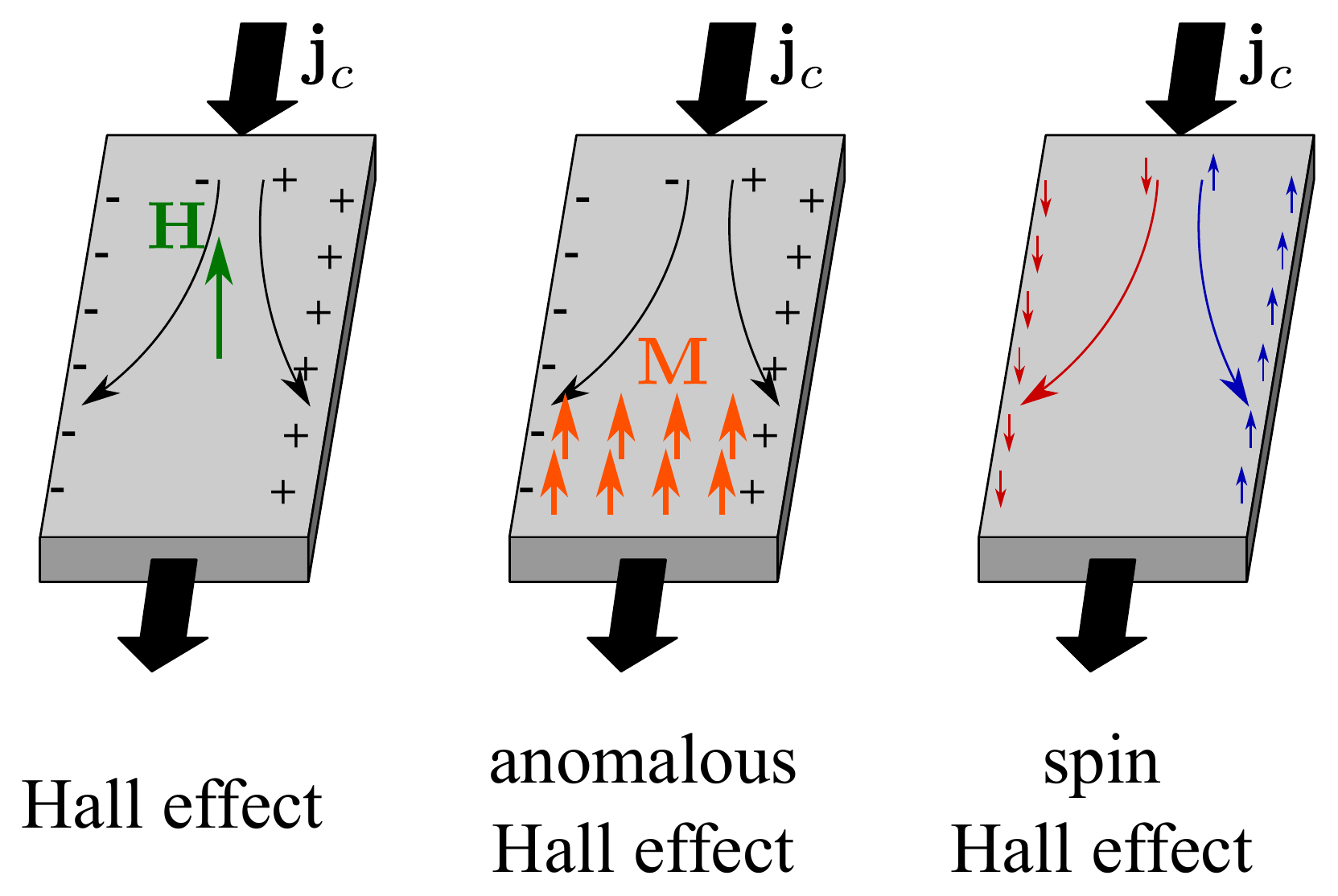}
\caption{Illustration of the spin Hall effect in comparison with the Hall effect and the anomalous
Hall effect. Here $\bj_c$ denotes the charge current, and $\bf M$ the magnetization.}
\label{she}
\end{center}
\end{figure}

\begin{figure}[b]
\begin{center}
\includegraphics[width=0.6\columnwidth]{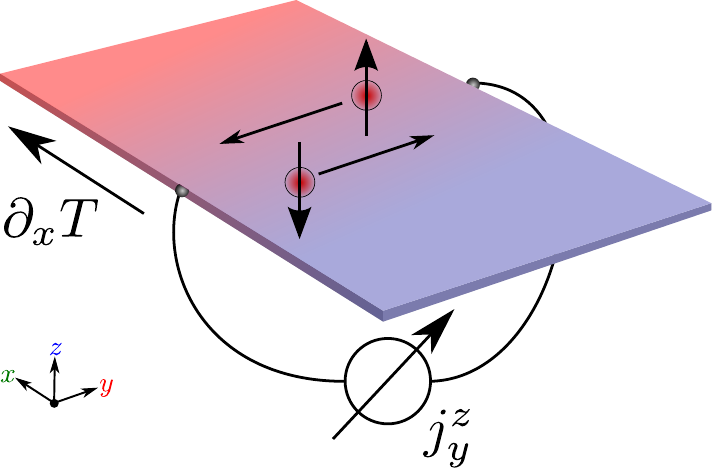}
\caption{Illustration of the spin Nernst effect, with $\partial_x T$ denoting the spatial gradient of
the temperature, and $j_y^z$ the spatial-y-component of the spin current with polarization in z-direction.}
\label{spin-nernst}
\end{center}
\end{figure}

Via spin-orbit coupling, an electric field or a temperature gradient applied to a two-dimensional
electron gas can generate both spin currents and spin polarisations, even in the absence of magnetic fields.
In particular, the generation of spin currents is known as spin Hall and spin Nernst effect,
respectively. In fact, ``spintronics'' and ``spin caloritronics'' have been a fast-growing field in
recent years, experimentally as well as theoretically.
We investigate these phenomena by means of a generalized Boltzmann equation which
takes into account spin-orbit coupling of both intrinsic and extrinsic origin \cite{gorini2010,raimondi2012}.
The spin Hall and Nernst effect are illustrated in Figs.\ \ref{she} and \ref{spin-nernst}.

For the thermally induced spin currents (transverse to the gradient) and polarisations,
the interplay between intrinsic and extrinsic mechanisms is shown to be critical. 
The relation between spin currents and spin polarisations is non-trivially affected by the
thermal gradient \cite{toelle2014}. It is argued that for
room-temperature experiments the $T$-dependence of electron-phonon scattering dominates over
scattering at (static) defects. For example, we find that the spin Hall conductivity is practically
independent of temperature for $T$ above the Debye temperature \cite{gorini2015}.
For details, see \cite{gorini2010,raimondi2012,toelle2014,gorini2015} and further references therein.
Our focus in this note is on two-dimensional systems, though some expression are easily generalized to
three dimensions \cite{toelle2014}.

In the next section, we briefly summarize the basic elements of the kinetic theory (Sec.\ \ref{kinetic}),
then we present in Sec.\ \ref{thermo} the results for the spin thermoelectric transport coefficients.
Section \ref{pss} is devoted to room-temperature phonon skew scattering. In the final section, Sec.\ \ref{sum},
we give a brief summary. 

\section{Kinetic theory}
\label{kinetic}

In order to set the stage, let us briefly discuss the standard model Hamiltonian for conduction electrons in a parabolic band \cite{shytov2006}:
\be
\label{model1}
\hat H_0^{\rm el} = \frac{p^2}{2m} - \frac{\alpha}{\hbar}\sigmav\times\hat{\bf z}\cdot\bp + V_{\rm imp}(\br) -
\frac{\lambda^2}{4\hbar}\sigmav\times\nabla V_{\rm imp}(\br)\cdot\bp \, .
\ee
The static lattice potential $V_{\rm crys}(\br)$ does not appear explicitly here, since its effects have been incorporated
in the effective mass $(m_0\rightarrow m)$ and effective Compton wavelength $(\lambda_0\rightarrow\lambda)$ \cite{winklerbook,Handbook}. 
Above, $\bhz$ is the unit vector pointing towards the metal-substrate interface, whereas $\bp, \br$ can be either
vectors in the $x$-$y$ plane for strictly 2D films, or also have a $z$-component for thicker, 3D systems. 
The second term on the r.h.s.\ is the Bychkov-Rashba \cite{bychkov1984} intrinsic spin-orbit coupling due to structure symmetry breaking,
characterized by the coupling constant $\alpha$. We recall that the intrinsic band splitting due to this term, denoted by $\Delta$, is
given by $2\alpha k_F$, where $k_F$ is the Fermi wavevector.
The random impurity potential $V_{\rm imp}(\br)$ enters directly and through the
fourth term, which represents the extrinsic spin-orbit interaction. In the strictly 2D limit the Hamiltonian \eqref{model1} was used to study the 
spin Hall \cite{hankiewicz2008,raimondi2009,raimondi2012,gorini2012} and Edelstein effect \cite{raimondi2012} in the presence of both intrinsic 
and extrinsic mechanisms at $T=0$.  Such mechanisms were shown {\it not} to be simply additive, and their interplay leads to a nontrivial 
behavior \cite{raimondi2009,raimondi2012}.

The free phonon part of the Hamiltonian, $H_0^{\rm ph}$, which is of the standard form \cite{agdbook,rammerbook1}, has to be added to \eqref{model1},
as well as the electron-phonon interaction. The latter, conveniently presented in second-quantized form, is given by (see, in particular, 
section 13 in \cite{agdbook}, and chapter 10.7 in \cite{rammerbook1})
\be
\label{model2}
\hat H^{\rm el-ph} = g \int \mathrm{d}\br \, \hat \varphi (\br) \hat\psi^\dag_\sigma (\br) \hat\psi_\sigma (\br) \, ,
\ee
where summation over the spin index $\sigma$ is implied. Correspondingly, we have to include $g \hat \varphi (\br)$ also in the 
(second-quantized form of the) last term on the r.h.s.\ of \eqref{model1}.

\begin{figure}
\includegraphics[width=0.8\columnwidth]{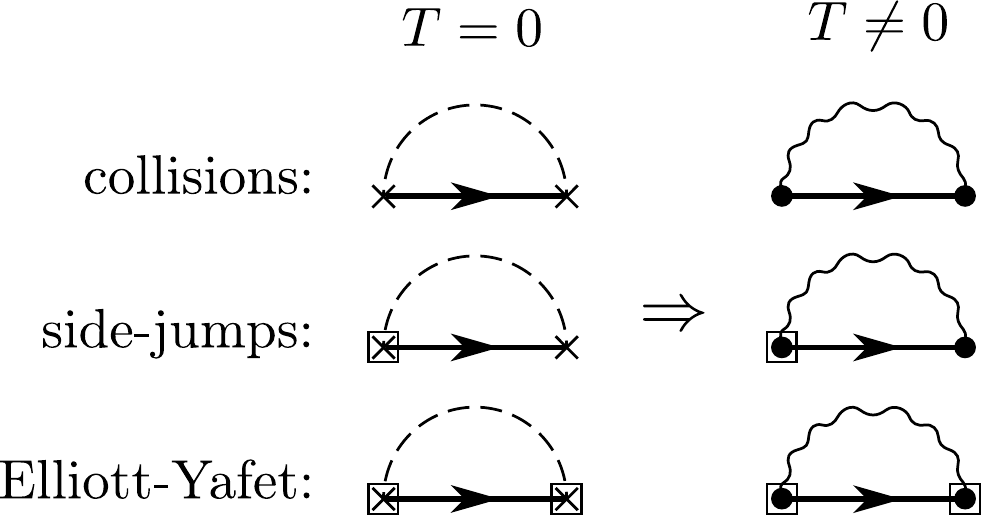}
\caption{Relevant self-energy contributions which determine the collision operators in the Boltzmann equation. The arrowed lines
represent the electron Green's function in Keldysh space, a cross (dot) the impurity (electron-phonon) vertex. The wavy line denotes
the phonon propagator, and a box around a vertex the spin-orbit coupling.}
\label{diagrams}
\end{figure}

The relevant self-energy diagrams are shown in Fig.\ \ref{diagrams}, l.h.s., for impurity scattering, and for
electron-phonon scattering, r.h.s. (For the moment, skew scattering is not considered; see Sec.\ \ref{pss}.)
It is to be expected that the latter dominates
for high temperature, i.e., for $T$ above the Debye temperature $T_D$. The kinetic (Boltzmann-like) equation for the $2\times2$ distribution 
function $f_{\bp}=f^0+\sigmav \cdot {\bf f}$, where $f^0$ and $\bf f$ are the charge and spin distribution functions, respectively,
reads \cite{gorini2010,raimondi2012,toelle2014}
\be 
\label{boltzmann}
\partial_t f_{\bp} + \tilde{\nabla} \cdot \left[ \frac{\bp}{m} f_{\bp} + \Delta {\bm j}_{\rm sj} \right] + 
\frac{1}{2} \left\lbrace{\Fcalv}\cdot{\nabla}_{\bp} , f_{\bp} \right\rbrace= I_{0}+I_{\rm sj}+I_{\rm EY} \, ,
\ee
where we introduced the covariant spatial derivative and the $SU(2)$ Lorentz force due to the Bychkov-Rashba spin-orbit coupling:
\ber
\tilde{\nabla}{}&{}={}&{}\nabla+\frac{i}{\hbar}\left[ \Acalv^a\frac{\sigma^a}{2},\cdot \right] \, ,\\
\Fcalv{}&{}={}&{}- \frac{\bp}{m} \times \boldsymbol{\mathcal{B}}^a \frac{\sigma^a}{2} \, ,\\
\mathcal{B}_i^a{}&{}={}&{}-\frac{1}{2\hbar}\varepsilon_{ijk} \varepsilon^{abc}\mathcal{A}_j^b\mathcal{A}_k^c \, .
\eer
A summation over identical indices is implied unless stated otherwise. Note that an external magnetic field is not included in these 
equations. The term $\Delta {\bm j}_{\rm sj}$ in \eqref{boltzmann} is a correction to the current due to side-jumps, given by
\be
\Delta {\bm j}_{\rm sj} = \frac{\lambda^2}{8\hbar \tau} \Big\langle \big\{ \left( \bp'-\bp \right) \times \sigmav , f_{\bp'} \big\} \Big\rangle_{\hat{\bp}'} \, ,
\label{SJcorr}
\ee
where $\langle \dots \rangle_{\hat{\bp}'}$ denotes the angular average.
The collision operators are not explicitly presented here. We only note that $I_0$ contains the standard terms describing momentum
relaxation due to electron-impurity and electron-phonon scattering; the total momentum relaxation rate is denoted by $1/\tau$.
For $T \gtrsim T_D$ the latter is similar to the former, since in
this limit electron-phonon scattering essentially is elastic, allowing for a simple addition of the corresponding rates 
(Matthiessen's rule). In the case of dominant electron-phonon scattering, the high-$T$ momentum relaxation rate therefore
is given by \cite{zimanbook2,rammerbook1}
\be
\frac{\hbar}{\tau} \simeq 2\pi \, (N_0 g^2) \, k_B T \; , \;\; T \gtrsim T_D \, ,
\ee
where $N_0$ denotes the density of states at the Fermi surface (per spin and volume). This high-$T$-expression is known to
be a good approximation even below $T_D$ (see, e.g., chapter IX, {\S} 5 in \cite{zimanbook2}). The last two terms on the r.h.s.\
of \eqref{boltzmann}, $I_{\rm sj}$ and $I_{\rm EY}$,
describe side-jump processes and Elliott-Yafet spin relaxation, respectively, cf.\ \cite{raimondi2012,toelle2014}; see also Fig.\ \ref{diagrams}.

From the distribution functions, the relevant physical quantities can be calculated. Here we present only the expressions
for the y-spin polarization and the z-polarized spin current flowing along the y-direction:
\be
s^y=\int \frac{\mathrm{d} \bp}{(2\pi\hbar)^2} f^y =  \int \mathrm{d} \epsilon_{\bp} N_0 \langle f^y \rangle \, ,
\label{spin}
\ee
\be
j_y^z=\mathrm{Tr} \, \frac{\sigma^z}{2} \int \frac{\mathrm{d} \bp}{(2\pi\hbar)^2}\left[\frac{p_y}{m} f_{\bp}+
    \frac{\lambda^2}{8\hbar\tau}\left\{ \left( \bp \times \sigmav \right)_y , f_{\bp} \right\} \right] \, .
\label{spin-current}
\ee
The second term on the r.h.s.\ of \eqref{spin-current} is due to side-jumps, cf.\ \eqref{boltzmann}. Due to the Bychkov-Rashba
term, a non-trivial relation between $s^y$ and $j_y^z$ is found:
\be
\partial_t s^y + \frac{2m\alpha}{\hbar} j_y^z = - \int \mathrm{d} \epsilon_{\bp} \frac{N_0}{\tau_{s}} \langle f^y \rangle \, .
\label{non-trivial}
\ee
Here we introduced the
Elliott-Yafet spin relaxation rate \cite{elliott1954,yafet63} (or spin flip rate, hence the subscript $s$):
\be
\frac{1}{\tau_{s}} = \frac{1}{\tau} \cdot \left(\frac{\lambda k}{2} \right)^4 \, ,
\ee
where $k \simeq k_F$. For an electric field the r.h.s.\ of \eqref{non-trivial} reduces to $-s^y/\tau_{s}$, 
but it does \emph{not} for a thermal gradient: in the latter case, the energy dependence of $1/\tau_s$ is found to be 
crucial \cite{toelle2014}.
The spin relaxation due to intrinsic spin-orbit coupling, named after Dyakonov and Perel \cite{dyakonov1971}, is
characterized by the following rate:
\be
\label{dp-rate}
\frac{1}{\tau_{\rm DP}} = \left( \frac{2m\alpha}{\hbar^2} \right)^2 D \, ,
\ee
where $D$ is the diffusion constant. This expression applies in the ``dirty'' limit, $\Delta \tau / \hbar \lesssim 1$.
More generally, $D$ has to be replaced by $D/[1 + (\Delta \tau /\hbar)^2]$.

Considering small variations of the temperature and a small electric field, the Boltzmann equation can be
linearized, and solved for the non-equilibrium part of the distribution function. Integral expressions for the 
transport coefficients follow; selected results are discussed in the following sections.

\section{Spin thermoelectric effects}
\label{thermo}

Efficient heat-to-spin conversion is the central goal of spin caloritronics \cite{bauer2012}.
When considering metallic systems, two phenomena stand out in this field:
the spin Nernst effect \cite{tauber2012,borge2013} and thermally-induced spin polarizations \cite{pang2010,dyrdal2013}.
They consist in the generation of, respectively, a spin current or a spin polarization
transverse to an applied temperature gradient. Note that in \cite{toelle2014}, and in this section, skew scattering is not 
taken into account. 

We note, in particular, that the spin Nernst conductivity $\sigma^{\rm sN}$ was recently investigated on the basis of ab initio methods
\cite{tauber2012}, predicting a linear $T$-dependence. 

As mentioned above, we consider the Boltzmann equation linearized in the temperature gradient and the electric field. Hence the
``drive term'' is proportional to
\be
 \partial_x f^{\rm eq}=  \left( \frac{\epsilon_{\bp}-\epsilon_F}{T} \partial_x T + 
 eE_x \right) \left( - \frac{\partial f^{\rm eq}}{\partial \epsilon_{\bp}} \right) \, ,
\ee
where $f^{\rm eq}$ is the Fermi function. The transport coefficients of interest are defined as follows:
\ber
    s^y {}&{} = {}&{} P_{\rm sE} E_x + P_{\rm sT} \partial_x T \, , \\
    j^z_y{}&{} = {}&{} \sigma_{\rm sE} E_x + \sigma_{\rm sT} \partial_x T \, .
\eer
Here we have chosen a symmetric notation with respect to the subscripts ``sE'' and ``sT''; of course, $\sigma_{\rm sE}$, usually denoted
as $\sigma^{\rm sH}$, is the spin Hall conductivity.

We obtain the following results (within the Sommerfeld expansion, $k_B T \ll \epsilon_F$; see \cite{toelle2014}): for the
Edelstein polarization coefficient we find
\be
P_{\rm sE} = -\frac{2m\alpha}{\hbar^2} \tau_{s} \cdot \sigma^{\rm sH} \, ,
\ee
while the spin Hall conductivity is given by
\be
\sigma^{\rm sH}=\frac{1}{1 + \tau_{s}/\tau_{\rm DP}} \left(\sigma_{\rm int}^{\rm sH}+\sigma_{\rm sj}^{\rm sH}\right) \, ;
\ee
furthermore,
\be
P_{\rm sT} = -S_0 \epsilon_F [{P_{\rm sE}}({\cal E})]^\prime_{\epsilon_F} \; ,
    \;\; \sigma_{\rm sT} = -S_0\epsilon_F [{\sigma^{\rm sH}}({\cal E})]^\prime_{\epsilon_F} \, .
\ee
Here $S_0 = - (\pi^2 k_B/3e) k_B T \, [\ln \sigma ({\cal E})]^\prime_{\epsilon_F}$ is the standard expression for the thermopower,
and the prime denotes differentiation with respect to energy; cf.\ chapter 7.9 in \cite{zimanbook1}.
In addition (see \cite{raimondi2012} and references therein):
\be
\label{sigma-int}
\sigma_{\rm int}^{\rm sH} = (e/4\pi\hbar)(\tau/\tau_{\rm DP}) = (e/2\pi\hbar^3) (\alpha k_F \tau )^2 \, ,
\ee
and \cite{Engel05,Tse06}
\be
\label{sigma-sj}
\sigma_{\rm sj}^{\rm sH} = e n \lambda^2 /4\hbar = e (\lambda k_F)^2 / 8\pi\hbar \; .
\ee
For the second equality in these equations, we used \eqref{dp-rate}, as well as $n = k_F^2 / 2\pi$ and $D = v_F^2 \tau / 2$.

\begin{figure}
\includegraphics[width=0.8\columnwidth]{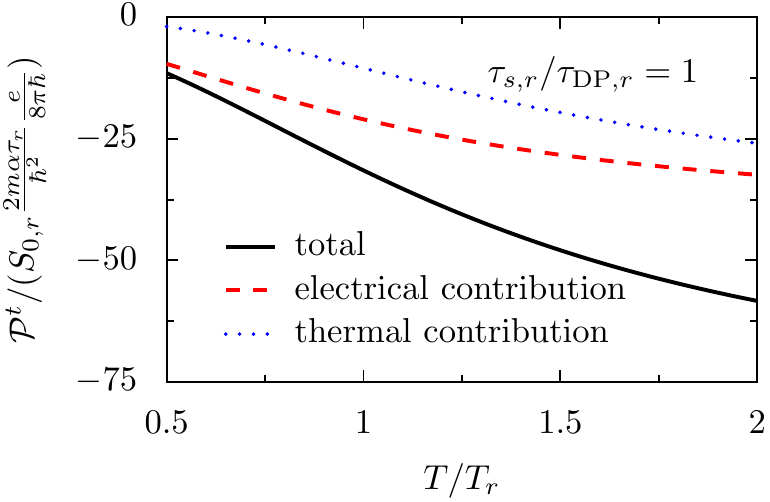}
\caption{Thermal Edelstein polarization coefficient, ${\cal P}^t$, versus $T/T_r$, in units of
$S_{0,r} (2m\alpha\tau_r/\hbar^2) (e/8\pi\hbar)$, split into its thermal and electrical contributions.
The Elliott-Yafet spin relaxation is chosen as $\tau/\tau_{s} = 0.01$; in addition, $\tau_{{s},r}/\tau_{{\rm DP},r} = 1$,
i.e., intrinsic and extrinsic spin relaxation are assumed to be of the same order of magnitude.
$T_r$ denotes the temperature scale (with the subscript $r$ referring to room temperature). Adapted from \cite{toelle2014}.}
\label{PtE_1}
\end{figure}

\begin{figure}
\includegraphics[width=0.8\columnwidth]{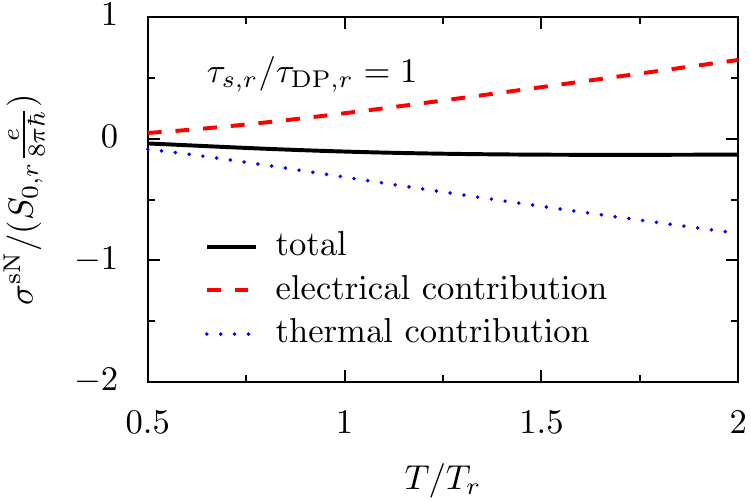}
\caption{Spin Nernst conductivity in units of the ``universal'' value of the intrinsic spin Hall conductivity times the Seebeck
coefficient at room temperature, $S_{0,r} \cdot e/8\pi\hbar$, versus $T/T_r$, for $\tau/\tau_{s} =0.01$ and
$\tau_{{s},r}/\tau_{{\rm DP},r} = 1$. $T_r$ denotes the temperature scale (with the subscript $r$ referring to room temperature).
Adapted from \cite{toelle2014}.}
\label{sigmasN_1}
\end{figure}

Finally, we consider concrete situations of experimental interest, namely (i) the thermal Edelstein effect, and
(ii) the spin Nernst effect, both for the case of open circuit conditions along the x-direction: this implies
$j_x=0$, and hence $E_x=S\partial_x T$, with the following results:
\be
\mathrm{(i)}: \quad s^y = \mathcal{P}^t \partial_x T \; , \;\; \mathcal{P}^t = S P_{\rm sE} + P_{\rm sT} \, ,
\ee
and
\be
\mathrm{(ii)}: \quad j_y^z = \sigma^{sN} \partial_x T \; , \;\; \sigma^{\rm sN}= S \sigma^{\rm sH} + \sigma_{\rm sT} \, .
\ee
In both cases, we can identify ``eletrical'' and ``thermal'' contributions. A non-linear $T$-dependence follows from
the fact that for high temperature $1/\tau \sim T$, hence $\tau_s \sim T^{-1}$ and $\tau_{\rm DP} \sim T$; see Figs.\
\ref{PtE_1} and \ref{sigmasN_1} for representative examples.

Summarizing this section, symmetric Mott-like formulas for the current or thermally induced spin polarization
(Edelstein effect, thermal Edelstein effect) and for the spin Hall and Nernst coefficients have been derived.  
The $T$-dependence of the transport coefficients is non-trivially affected by the competition between intrinsic 
and extrinsic spin-orbit coupling mechanisms, $\tau_{\rm DP}$ versus $\tau_{s}$.  
In the diffusive regime the relaxation times have different $T$-dependences, which ultimately causes
a non-linear $T$-behavior. The non-linearity is in general stronger for the thermal Edelstein effect, and, especially 
in the spin Nernst case, it becomes weaker with decreasing intrinsic spin-orbit coupling strength.

\section{Phonon skew scattering}
\label{pss}

\begin{figure}
\includegraphics[width=0.4\columnwidth]{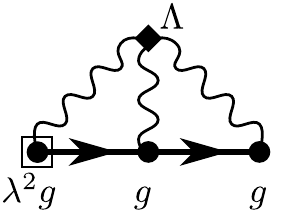}
\caption{Phonon skew scattering self-energy; $g$ denotes the electron-phonon
vertex, and $\Lambda$ the parameter describing the strength of the anharmonic lattice contribution, i.e., 3-phonon
processes. Note that $\Lambda$ is proportional to the negative of the Gr\"uneisen parameter $\gamma$, namely
$\Lambda = -\gamma / (\rho^{1/2} v_s)$, where $\rho$ and $v_s$ are the ionic mass density and the sound velocity,
respectively.}
\label{diagram-pss}
\end{figure}

A diversity of spin Hall effects in metallic systems is known to rely on Mott skew scattering.
In this section its high-$T$ counterpart, phonon skew scattering (pss), is investigated. One of the corresponding
self-energy diagrams is shown in Fig.\ \ref{diagram-pss}. As a central result, the pss spin Hall conductivity is found 
to be practically $T$-independent for $T$ above the Debye temperature $T_D$ \cite{gorini2015}.  

As discussed in detail in \cite{gorini2015}, a certain $T=0 \rightarrow T>T_D$ correspondence lets
us immediately turn known $T=0$ results into their $T>T_D$ counterparts, with the result that
the full expression for the high-$T$ spin Hall conductivity
is structurally similar to the $T=0$ expressions appearing in \cite{raimondi2012}.
Explicitly, for a 2D homogeneous bulk system: 
\be
\label{sHfull}
\sigma^{\rm sH} = \frac{1}{1+\tau_{s}/\tau_{\rm DP}}\left(\sigma^{\rm sH}_{\rm int} 
+ \sigma^{\rm sH}_{\rm sj} + \sigma^{\rm sH}_{\rm ss}\right) \, ,
\ee
where the intrinsic part of the spin Hall conductivity and the side-jump contribution were introduced in the previous
section, see \eqref{sigma-int} and \eqref{sigma-sj}. In addition, for $T=0$ where phonons can be neglected, 
one finds \cite{Engel05,Tse06}
\be
\label{ss0}
\sigma_{\rm ss, 0}^{\rm sH} = 2\pi \left(\frac{\lambda k_F}{4}\right)^2\frac{en}{m} \,  N_0 v_0 \tau \, ,
\ee
with $N_0=m/2\pi\hbar^2$, and 
$v_0$ the scattering amplitude. Here $\tau$ is the {\em impurity} scattering time. We emphasize
that the side-jump spin Hall conductivity is independent of 
the scattering mechanism (at least in simple parabolic bands), whereas the skew scattering contribution
is proportional to $\tau$, i.e., to the Drude conductivity $\sigma_D = e^2n\tau/m = en\mu$ ($e>0$).

Considering the phonon skew scattering self-energy in detail, a major simplification arises in the high-$T$ limit,
where the ``greater'' and ``lesser'' phonon Green's functions can be approximated by their classical limits. Hence
we find that the pss self-energy has the standard form due to the coupling to an external field, whose role is played
here by a quantity which we denoted by $\mathbb{D}$ \cite{gorini2015}.
Exploiting the fact that the phonon energies ($\sim \hbar\omega_D$) are small compared to 
$\hbar\omega \sim k_B T$, one finds
\be
\mathbb{D}_{123} \approx -3\Lambda g^3 (k_B T)^2 \, .
\ee
The $T=0\rightarrow T>T_D$ correspondence for skew scattering thus explicitly reads
\be
\label{correspondence_3}
n_i v_0^3 \rightarrow -3\Lambda g^3 (k_B T)^2 \, ,
\ee
where $n_i$ is the density of impurities. This yields at once \cite{raimondi2012,gorini2015}
\be
\label{ssT3}
\sigma_{\rm ss}^{\rm sH} = -3\left(\frac{\lambda k_F}{4}\right)^2 \frac{en}{m} \frac{\hbar \Lambda}{g} \, .
\ee
This shows that the pss spin Hall conductivity at high temperature is $T$-independent, in particular, it does {\em not} scale
as the mobility (which was suggested in earlier works \cite{hankiewicz2006,vila2007,vignale2010,niimi2011,isasa2014}, based
on the $T=0$ expressions). Note that $\Lambda < 0$ \cite{zimanbook2}.

\begin{figure}[t]
\includegraphics[width=0.8\columnwidth]{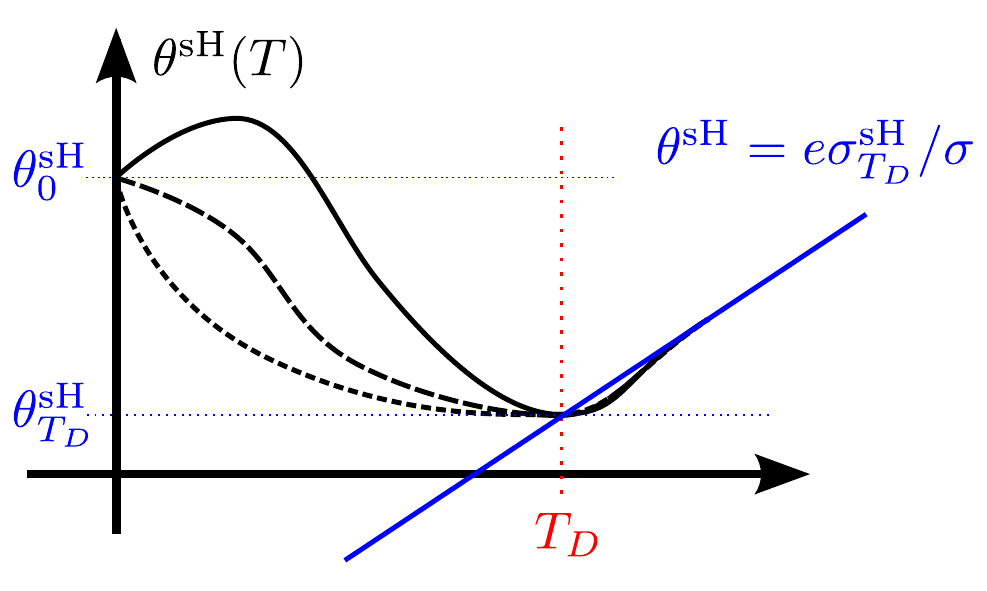}
\caption{Conjectured temperature dependences of the spin Hall angle, $\theta^{\rm sH} \equiv e\sigma^{\rm sH} / \sigma$, based
on the assumption $\theta^{\rm sH}_0 > \theta^{\rm sH}_{T\sim T_D} > 0$ together with the result \eqref{theta-high}.}
\label{conjecture}
\end{figure}

Bearing in mind the relations $j_y^z = \sigma^{\rm sH} E_x$ and $j_x = \sigma E_x$ for the spin and the charge current,
respectively, the (dimensionless) spin Hall angle is defined by $\theta^{\rm sH} = e j_y^z / j_x$, i.e., 
$\theta^{\rm sH} = e \sigma^{\rm sH} / \sigma$. For an estimate, consider first $T=0$ and the dirty limit
(see above) $\alpha k_F \tau / \hbar < 1$. Dropping all numerical factors and $\hbar$'s, we find:
\be
\theta^{\rm sH}_0 \sim \frac{1/\tau}{\epsilon_F} \cdot
\frac{(\alpha k_F \tau)^2 + (\lambda k_F)^2 + (\lambda k_F)^2 (\epsilon_F \tau) (N_0 v_0)}{1 + (\alpha\tau /\lambda)^2 / (\lambda k_F)^2} \, ,
\ee
where the three terms in the numerator correspond to the three terms displayed on the r.h.s.\ of \eqref{sHfull}. Since
$N_0 v_0 \sim 1/2$, it appears that the skew scattering term dominates over the side-jump contribution. (Note, however, that
$v_0$ can be of either sign; for the following discussion, we assume $v_0 > 0$.) Obviously, a more quantitative
estimate is difficult since the parameters are material-dependent and generally not precisely known. In order to proceed, let us
assume that intrinsic spin-orbit coupling is small, $\alpha < \lambda^2 k_F / \tau$. Then we may neglect the corresponding terms in
the numerator and the denominator, with the result
\be
\theta^{\rm sH}_0 \sim (\lambda k_F)^2 \cdot N_0 v_0 \, .
\ee
Since $\tau$ decreases with increasing temperature, this approximation improves with increasing $T$,
and we find from \eqref{ssT3} the following estimate:
\be
\label{theta-high}
\theta^{\rm sH}_T \sim - (\lambda k_F)^2 \cdot \Lambda / (g \tau) \sim T \, .
\ee
In particular, we realize that $\theta^{\rm sH}_0 > \theta^{\rm sH}_{T\sim T_D}$. Thus the $T$-dependence of the spin Hall
angle is non-monotonous, see Fig.\ \ref{conjecture}.

\section{Discussion}
\label{sum}

In our recent works \cite{toelle2014,gorini2015}, we have been able to extend the kinetic theory of spin-orbit coupled
electron (or hole) systems to finite (room) temperature, where momentum relaxation is dominated (in most cases) by
electron-phonon scattering. The calculations are simplified in the high-$T$ limit, $T > T_D$, where electron-phonon processes
are elastic. This is particularly useful for the phonon skew scattering contribution to the spin Hall conductivity. We
conjecture that the $T$-dependence of the spin Hall angle, for weak intrinsic spin-orbit coupling, may become non-monotonous.

\acknowledgments{We are grateful to Gurucharan Vijay Karnad and Mathias Kl\"aui for stimulating discussions.
Financial support from the Deutsche Forschungsgemeinschaft through SFB 698 and TRR 80 is gratefully acknowledged.}

\bibliography{biblio}

\end{document}